\documentclass[twoside,twocolumn,aps,prc,superscriptaddress,nofootinbib,showpacs]{revtex4}
\usepackage{amsmath,amssymb,graphicx}

\usepackage{multirow}

\usepackage{amsfonts}
\usepackage{dsfont}

\makeatletter

\newcommand{\Rmnum}[1]{\expandafter\@slowromancap\romannumeral #1@}
\makeatother

\allowdisplaybreaks

\begin{document}

\title{$Z_b(10650)$ and $Z_b(10610)$ states in a chiral quark model}

\author{M.T. Li}
 \affiliation{Institute of High Energy Physics, CAS, P.O. Box 918-4, Beijing 100049, China}
 \affiliation{Theoretical Physics Center for Science Facilities (TPCSF), CAS, Beijing 100049, China}

\author{W.L. Wang}
 \affiliation{Institute of High Energy Physics, CAS, P.O. Box 918-4, Beijing 100049, China}
 \affiliation{Theoretical Physics Center for Science Facilities (TPCSF), CAS, Beijing 100049, China}

\author{Y.B. Dong}
 \affiliation{Institute of High Energy Physics, CAS, P.O. Box 918-4, Beijing 100049, China}
 \affiliation{Theoretical Physics Center for Science Facilities (TPCSF), CAS, Beijing 100049, China}

\author{Z.Y. Zhang}
 \affiliation{Institute of High Energy Physics, CAS, P.O. Box 918-4, Beijing 100049, China}
 \affiliation{Theoretical Physics Center for Science Facilities (TPCSF), CAS, Beijing 100049, China}

\begin{abstract}

We perform a systematic study of $B\bar{B}^*$, $B^*\bar{B}^*$, $D\bar{D}^*$ and $D^*\bar{D}^*$
systems by using effective interaction in our chiral quark model. Our results show that
the interactions of $B\bar{B}^*$, $B^*\bar{B}^*$, $D\bar{D}^*$ and $D^*\bar{D}^*$ states
are attractive, which consequently result in $B\bar{B}^*$, $B^*\bar{B}^*$, $D\bar{D}^*$
and $D^*\bar{D}^*$ bound states. The recent observed exotic-like hadrons of $Z_b(10610)$ and
$Z_b(10650)$ are, therefore in our approach, interpreted as loosely  bound states of $B\bar{B}^*$
and $B^*\bar{B}^*$, while $X(3872)$ and $Y(3940)$ are $D\bar{D}^*$ and $D^*\bar{D}^*$ molecule
states, respectively.

\end{abstract}



\maketitle

\section{Introduction}

Recently, the Belle Collaboration pronounced two charged bottomonium-like
resonances $Z_b$(10610) and $Z_b$(10650) \cite{belle0}. These two resonances
were observed in the five decay channels of $\Upsilon(5S)\rightarrow\pi^\pm\Upsilon(nS)$
($n=1,2,3$) and $\Upsilon(5S)\rightarrow\pi^\pm h_b(mP)$ ($m=1,2$). The recommended
masses and widths of the two resonances are $M(10610)=10608.4\pm 2.0$ MeV/$c^2$,
$\Gamma_1=15.6\pm 2.5$ MeV/$c^2$ and $M(10650)=10653.2\pm1.5$ MeV/$c^2$,
$\Gamma_2=14.4\pm3.2$ MeV/$c^2$, respectively. For both states, experimental results
fixed their quantum numbers as $I^{G}(J^{P})$=$1^{+}(1^{+})$. The two resonances
can't be simply interpreted as the $b\bar{b}$ heavy quarkoniums because they are charged.
It is suggested that they might be $B\bar{B}^*$ and $B^*\bar{B}^*$ molecules, respectively,
since their masses are near to the thresholds of $B\bar{B}^*$ and $B^*\bar{B}^*$, and they
have narrow widths.

There are many different explanations about the structures of $Z_b$(10610)and $Z_b$(10650)
 in the literature \cite{ali1,ali2,tguo,cui,matheus,wchen,huang,nieves,dong0,ding1,ding2,wong,he,guo}.
Ali et al. \cite{ali1,ali2} and Guo et al. \cite{tguo} supported that $Z_b$(10610) and $Z_b$(10650)
could be explained as tetraquark states. Cui et al. \cite{cui} showed both tetraquark and molecule
interpretations were compatible with the experimental measurements in the QCD sum rule calculation.
However, Matheus \cite{matheus} and Chen \cite{wchen} argued that the masses of tetraquark states
were obviously below the $B\bar{B}^*$ and $B^*\bar{B}^*$ thresholds. Zhang et al.\cite{huang} showed
their quite satisfactory result that $Z_b(10610)$ might be regarded as a $B\bar{B}^*$ molecule which
agreed with the Belle's molecule suggestion \cite{belle0}. Moreover, Nieves and Valderrama argued
that there might be three isoscalar $B\bar{B}^*$ bound states by considering one pion exchange interaction,
but all of the three didn't fit the quantum numbers of $Z_b(10610)$ \cite{nieves}. Ding et al.\cite{ding1,ding2}
suggested $B\bar{B}^*$ and $B^*\bar{B}^*$ could form molecule bound states, but these states
couldn't fit to the $1^{+}(1^{+})$ quantum numbers of the two $Z_b$ states. By including only screened
color-Coulomb and screened linear confining potentials, Wong \cite{wong} also got some $B\bar{B}^*$
and $B^*\bar{B}^*$ molecules, but his $Z_b$(10610) and $Z_b$(10650) should be the first radial excitation
of $B\bar{B}^*$ and $B^*\bar{B}^*$ molecules, respectively. In addition, by considering both one boson
exchange interaction and the one pion exchange interaction and considering the $S-D$ mixing
between $B/B^{*}$ and $\bar{B}^*$, Sun et al. \cite{he} gave their result of $Z_b$(10610) and $Z_b$(10650) in
the molecule scenario. It should be mentioned that the calculations of Dong et al. \cite{dong0} and
Cleven et al. \cite {guo} supported the molecule scenario, however, the latter couldn't
exclude the tetraquark component.

We see that by using different approaches one may have different interpretations for the two new resonances
of $Z_b(10650)$ and $Z_b(10610)$. In this work, we will study the two resonances based on our chiral quark model.
It should be stressed that the chiral quark model is successful to well describe the binding energies of the
baryon ground states, the binding energy of the deuteron, the $NN$ scattering phase shifts and the $NY$ cross
sections simultaneously \cite{zhang1,zhang2,dai,liu1,liu2,wangwlsigc,wang2007,wang2008,wang2010,wang2011}.
It is expected that the chiral quark model provides another way to study the possible $B\bar{B}^*$ and $B^*\bar{B}^*$
bound states problem on quark level.

In addition, it is known that the study of the hadronic states of $X(3872)$ observed in Ref. \cite{belle1, cdf, d0, barbar0}
and Y(3940) observed in Ref. \cite{belle2005,belle2006,belle2007,barbar2008} have been a hot issue recently.
This is because they have exotic properties and can't be simply regarded as $c\bar{c}$ charmoniums. Among the
various approaches, hadronic molecule of $0^+(1^{++})$ $D\bar{D}^*$ is one feasible interpretation of $X(3872)$
\cite{swanson1, swanson2, dong, lee, liu3, liux}. Therefore, analogous to the systems of $B\bar{B}^*$ and
$B^*\bar{B}^*$, it is also of significance to study the $D\bar{D}^*$ and $D^*\bar{D}^*$ systems in our
chiral quark model for $X(3872)$ and Y(3940).

It should be mentioned that several works based on our chiral quark
model have been published \cite{liu1, liu2, wangwlsigc} to study the
bound states and binding energies of heavy-meson molecule systems
with Resonating Group Method (RGM). The RGM method is a successful
one proficiently adopted in Ref. \cite{zhang1, zhang2, dai, liu1,
liu2, wangwlsigc, wang2007, wang2008, wang2010, wang2011}. In this
work, different from the RGM method, we apply the approach
introduced in Ref. \cite{li} to derive the analytical from of the
total interaction potentials between the two S-wave heavy mesons.
Here the interaction potentials result from the one meson exchange
interactions between two light quarks. We anticipate that this
method would give a more accurate description of the short-range
interaction between the two clusters in the chiral quark model, and
we thoroughly investigate the possible bound states of $B\bar{B}^*$
and $B^*\bar{B}^*$ systems as well as their charm partners
$D\bar{D}^*$ and $D^*\bar{D}^*$ by solving the Schr\"{o}dinger
Equation with our analytical interaction potentials between the two
clusters.

The paper is organized as follows. In section \ref{sec:formulism}, the framework of our chiral quark model
is briefly introduced, and the analytical forms of the effective interaction potentials between the two S-wave meson clusters
under our chiral quark model are given. The bound state solutions for the $B\bar{B}^*$ and $B^*\bar{B}^*$ systems
and their charm partners of  $D\bar{D}^*$ and $D^*\bar{D}^*$ are shown in Sec.~\ref{sec:result}.
Finally, the summary is given in Sec.~\ref{sec:sum}.

\section{Formulation} \label{sec:formulism}

The framework of our effective chiral quark model has been discussed extensively in the
literature \cite{zhang1,zhang2,dai,liu1,liu2}. Here we just briefly present the main part
of the model. The total Hamiltonian of the approach reads
\begin{eqnarray}
H=T_q-T_{G}+\sum_{i,j}V(r_{ij}),
\end{eqnarray}
where $T_q$ is the sum of the kinetic energy operator of all the quarks and antiquarks in the two clusters,
and $T_G$ is the kinetic energy operator of the center-of-mass motion. In a molecule
scenario or cluster model, because the interactions inside each cluster don't dedicate
to the total interaction between the two clusters of the system, only the interactions
between the two different clusters need to be taken into account. $V(r_{ij})$ in Eq. 1
represents the interactions between the $i$th light quark or antiquark in the first cluster and
the $j$th light quark or antiquark in another one, and

\begin{eqnarray}
V(r_{ij})=V^{OGE}(r_{ij})+V^{conf}(r_{ij})+V^M(r_{ij}).
\end{eqnarray}
Here $V_{ij}^{OGE}$ is the one-gluon exchange (OGE) potential

\begin{eqnarray} \label{eq:oge}
V^{OGE}(r_{ij}) &=& \frac{1}{4}g_i g_j\left(\lambda^c_i\cdot\lambda^c_j\right) \Bigg[\frac{1}{r_{ij}} -\frac{\pi}{2} \delta(r_{ij}) \nonumber \\
&& \times\left(\frac{1}{m^2_{q_i}}+\frac{1}{m^2_{q_j}}+\frac{4}{3}\frac{\sigma_i\cdot\sigma_j}{m_{q_i}m_{q_j}} \right)\Bigg],
\end{eqnarray}
and $V_{ij}^{conf}$ is the confinement

\begin{eqnarray} \label{eq:conf}
V^{conf}(r_{ij})=-({\lambda}_{i}^{c}\cdot{\lambda}_{j}^{c})\left(a_{ij}^{c}r_{ij} +a_{ij}^{c0}\right).
\end{eqnarray}

$V^{M}(r_{ij})$ in Eq. 2 stands for the total one-meson exchange interaction potential. Generally,
\begin{eqnarray}
V^{M}(r_{ij})=\sum^8_{a=0} V^{\sigma_a}(r_{ij})+\sum^8_{a=0} V^{\pi_a}(r_{ij})+\sum^8_{a=0} V^{\rho_a}(r_{ij}),
\end{eqnarray}
with $V^{\sigma_a}(r_{ij})$, $V^{\pi_a}(r_{ij})$ and $V^{\rho_a}(r_{ij})$ being the interactions induced
from scalar meson, pseudoscalar meson, and vector meson exchanges, respectively.

For light quark-quark interaction, $V^{\sigma_a}(r_{ij})$, $V^{\pi_a}(r_{ij})$ and $V^{\rho_a}(r_{ij})$ have such forms:
\begin{eqnarray}
V^{\sigma_a}(r_{ij})&=& -C(g_{ch},m_{\sigma_a},\Lambda) X_1(m_{\sigma_a},\Lambda,r_{ij})
\left(\lambda^a_i\lambda^a_j\right),\label{scalar} \\
V^{\pi_a}(r_{ij})&=& C(g_{ch},m_{\pi_a},\Lambda) \frac{m^2_{\pi_a}}{12m_im_j} X_2(m_{\pi_a},\Lambda,r_{ij}) \nonumber \\
&& \times \left(\sigma_i\cdot\sigma_j\right)\left(\lambda^a_i\lambda^a_j\right), \label{pseudoscalar}\\
V^{\rho_a}(r_{ij}) &=& C(g_{\rm chv},m_{\rho_a},\Lambda)\Bigg[X_1(m_{\rho_a},\Lambda,r_{ij}) + \frac{m^2_{\rho_a}}{6m_im_j} \nonumber \\
&& \times \left(1+\frac{f_{\rm chv}}{g_{\rm chv}} \frac{m_i+m_j}{M_N}+\frac{f^2_{chv}}{g^2_{chv}}
\frac{m_im_j}{M^2_N}\right)   \nonumber \\
&& \times \,  X_2(m_{\rho_a},\Lambda,r_{ij}) \, (\sigma_i\cdot\sigma_j)\Bigg] \left(\lambda^a_i\lambda^a_j\right),\label{vector}
\end{eqnarray}
where $\lambda^a$ is the Gell-Mann matrix in flavor space. In our chiral SU(3) quark model,
we only consider the scalar meson and pseudoscalar meson exchanges, while in our extended chiral SU(3) quark model
vector meson exchange interactions are also included. $m_{\sigma_a}$, $m_{\pi_a}$ and
$m_{\rho_a}$  in Eq. 5 are the masses of the scalar nonets, the pseudoscalar nonets
and the vector nonets, respectively. $M_N$ in Eq. 8 is a mass scale usually taken as the mass of
nucleon \cite{dai}. In the above equations, $m_i$ is the $i$th light constituent quark mass in the first cluster,
$m_j$ is the $j$th light constituent quark mass in the second cluster.
$g_{ch}$ is the coupling constants for the scalar and pseudoscalar nonets.
$g_{chv}$ and $f_{chv}$ are the coupling constants for the
vector coupling and tensor coupling of vector nonets, respectively.
In Eqs. \ref{scalar}, \ref{pseudoscalar}, \ref{vector},

\begin{eqnarray}
C(g_{ch},m,\Lambda) &=& \frac{g^2_{ch}}{4\pi} \frac{\Lambda^2}{\Lambda^2-m^2} m, \\
\label{ev1} X_1(m,\Lambda,r_{ij}) &=& Y(mr_{ij})-\frac{\Lambda}{m} Y(\Lambda r_{ij}), \\
\label{ev2} X_2(m,\Lambda,r_{ij}) &=& Y(mr_{ij})-\left(\frac{\Lambda}{m}\right)^3 Y(\Lambda r_{ij}), \\
Y(x) &=& \frac{1}{x}e^{-x}.
\end{eqnarray}

For light quark-antiquark interactions, $G_{\sigma_a,\pi_a,\rho_a}$, the $G$-parity of the
exchanged meson, should be included in the interaction potentials $V^{\sigma_a}(r_{ij})$,
$V^{\pi_a}(r_{ij})$ and $V^{\rho_a}(r_{ij})$.

In the $B\bar{B}^*$, $B^*\bar{B}^*$, $D\bar{D}^*$ and $D^*\bar{D}^*$ two-heavy-meson systems,
only the light quark-antiquark interactions are considered. To solve the bound state problem
for these systems, unlike the Resonating Group Method (RGM) employed in our previous works
\cite{liu1,liu2}, we use the approach explicitly discussed in Ref. \cite{li}. We first derive
an analytical form of the total effective interaction between the two mesons, and then by
solving the Schr\"{o}dinger Equation, we calculate the binding energy $E$,
and finally draw a conclusion whether the molecule bound states could exist in the
two-meson systems or not.

Now corresponding to scalar, pseudoscalar and vector meson exchanges (see Eqs. \ref{scalar}, \ref{pseudoscalar}
and \ref{vector}), using the method described in Ref.  \cite{li}, we get the analytical effective
interaction potentials between the two S-wave heavy meson clusters as

\begin{eqnarray}
V_{q\bar{q}}^{\sigma_a}(\xi)&=& -G_{\sigma_a}C(g_{ch},m_{\sigma_a},\Lambda) X_{1q\bar{q}}(m_{\sigma_a},\Lambda,\xi)\left(\lambda^a_q\lambda^a_{\bar{q}}\right),\label{eff:scalar} \\
V_{q\bar{q}}^{\pi_a}(\xi)&=& G_{\pi_a}C(g_{ch},m_{\pi_a},\Lambda) \frac{m^2_{\pi_a}}{12m_qm_{\bar{q}}} X_{2q\bar{q}}(m_{\pi_a},\Lambda,\xi) \nonumber \\
&& \times \left(\sigma_q\cdot\sigma_{\bar{q}}\right)\left(\lambda^a_q\lambda^a_{\bar{q}}\right), \label{eff:pseudoscalar}\\
V_{q\bar{q}}^{\rho_a}(\xi) &=& G_{\rho_a}C(g_{chv},m_{\rho_a},\Lambda)\Bigg[X_{1q\bar{q}}(m_{\rho_a},\Lambda,\xi)  \nonumber \\
&&+ \frac{m^2_{\rho_a}}{6m_qm_{\bar{q}}}\left(1+\frac{f_{chv}}{g_{chv}} \frac{m_q+m_{\bar{q}}}{M_N}+\frac{f^2_{chv}}{g^2_{chv}}  \frac{m_qm_{\bar{q}}}{M^2_N}\right)   \nonumber \\
&& \times X_{2q\bar{q}}(m_{\rho_a},\Lambda,\xi) \, (\sigma_q\cdot\sigma_{\bar{q}})\Bigg] \left(\lambda^a_q\lambda^a_{\bar{q}}\right).\label{eff:vector}
\end{eqnarray}
Here, $\xi$ is the relative coordinate between two different clusters, namely, the relative
coordinate between the two centers-of-mass of the two clusters, and
\begin{eqnarray}
\label{eff:ev1} X_{1q\bar{q}}(m,\Lambda,\xi) &=& Y_{q\bar{q}}(m\xi)-\frac{\Lambda}{m} Y_{q\bar{q}}(\Lambda \xi), \\
\label{eff:ev2} X_{2q\bar{q}}(m,\Lambda,\xi) &=& Y_{q\bar{q}}(m\xi)-\left(\frac{\Lambda}{m}\right)^3 Y_{q\bar{q}}(\Lambda \xi).
\end{eqnarray}
The modified Yukawa term reads
\begin{eqnarray}
Y_{q\bar{q}}(m\xi)&=&\frac{1}{2m\xi}e^{\frac{m^2}{4\beta}}\bigg\{e^{-m\xi}
\Big\{1-erf\Big[-\sqrt{\beta}(\xi-\frac{m}{2\beta})\Big]\Big\}\nonumber \\
&&-e^{m\xi}\Big\{1-erf\Big[\sqrt{\beta}(\xi+\frac{m}{2\beta})\Big]\Big\}\bigg\}.\label{modified Yukawa}
\end{eqnarray}
Here,
\begin{eqnarray}
\beta&=&\frac{\mu_{q\bar{Q}}\mu_{Q\bar{q}}\omega}{\mu_{q\bar{Q}}
\left(\frac{m_{Q}}{m_Q+m_{\bar{q}}} \right)^2+\mu_{Q\bar{q}} \left(\frac{m_{\bar{Q}}}{m_q+m_{\bar{Q}}} \right)^2},
\end{eqnarray}
and $\omega$ is the harmonic-oscillator frequency of the light quark wave function. $q$ ($\bar{q}$) and $Q$
($\bar{Q}$) in Eq. 19 are the light quark (antiquark) and heavy quark (antiquark)in the two clusters, respectively.

There are some necessary parameters in the interaction potentials in our chiral quark model.
In this work, we adopt the parameters determined in our previous works \cite{dai,liu1,liu2,zhang1,zhang2,wang2007,
wang2008, wang2010}. The up/down quark mass $m_q$ is taken as $313$  MeV. The harmonic-oscillator frequency
$\omega$ is $2.522 fm^{-1}$  in the chiral SU(3) quark model and $3.113 fm^{-1}$ in the extended chiral
SU(3) quark model. The coupling constant for the scalar and pseudoscalar chiral field $g_{ch}=2.621$
is fixed by the relation of
\begin{eqnarray}
\frac{g^{2}_{ch}}{4\pi} =\frac{9}{25} \frac{g^{2}_{NN\pi}}{4\pi} \frac{m^{2}_{u}}{M^{2}_{N}},\nonumber
\end{eqnarray}
with $g^{2}_{NN\pi}/4\pi=13.67$ determined from experiments. In our extended chiral SU(3) quark model, vector
meson exchange interactions are included. The coupling constant for vector coupling $g_{chv}$ in
Eq. \ref{vector} is taken to be $2.351$ and $1.973$, respectively. Correspondingly, the ratio of
the tensor coupling $f_{chv}$ to the vector coupling $g_{chv}$ is taken to be 0 and 2/3,
respectively. In our calculation, the masses of the mesons are taken from the PDG \cite{PDG2010}, except
the $\sigma$ meson, which does not have a well-defined value. Here $m_\sigma$ is obtained by fitting the
binding energy of the deuteron \cite{dai}. It is $m_\sigma=595$  MeV in
our chiral SU(3) quark model, 535  MeV and 547  MeV for $f_{chv}/g_{chv}$ respectively taken as 0 and 2/3
in the extended chiral SU(3) quark model. The cutoff $\Lambda$ is taken as $1100$   MeV. Since there is no
color-interrelated interaction between the two color-singlet clusters, we don't have to consider the OGE
and confinement interactions, and therefore, we don't list
the OGE coupling constant $g_{i,j}$ in Eq. \ref{eq:oge} and the parameters of the confinement
potential $a_{ij}^{c}$ and $a_{ij}^{c0}$ in Eq. \ref{eq:conf}.

The remaining parameters to be determined are the heavy quark masses $m_c$ and $m_b$.
In our work, we find that the final results are not sensitive to the variation of the heavy quark masses,
and we take  $m_c=1430$  MeV \cite{zhanghx1} and $m_b=4720$  MeV \cite{zhanghx2} as typical values.

\section{NUMERICAL SOLUTIONS}\label{sec:result}

To make sure the $B\bar{B}^*, B^*\bar{B}^*$ systems and their charm partners of  $D\bar{D}^*, D^*\bar{D}^*$
have definite quantum numbers of isospin $I$ and $C$-parity $C$, we construct the following flavor wave
functions as Refs. \cite{liu1,liux,he}. For $B\bar{B}^*$ system:
\begin{eqnarray}
&I=1:&\begin{cases}
\dfrac{1}{\sqrt{2}}(B^{*+}\bar{B}^0+cB^{+}\bar{B}^{*0})\\
\dfrac{1}{\sqrt{2}}(B^{*-}\bar{B}^0+cB^{-}\bar{B}^{*0})\\
\dfrac{1}{2}[(B^{*0}\bar{B}^{0}-B^{\ast+}B^{-})+c(B^{0}\bar{B}^{*0}-B^{+}B^{*-})],
\end{cases} \\
&I=0:&\frac{1}{2}[(B^{*0}\bar{B}^{0}+B^{*+}B^{-})+c(B^{0}\bar{B}^{*0}+B^{+}B^{*-})].
\end{eqnarray}
Here, $c$=1 for $C$=+ and $c$=--1 for $C$=--. For $B^*\bar{B}^*$ system, the flavor function is
\begin{eqnarray}
&I=1:&\begin{cases}
B^{*+}\bar{B}^{*0}\\
B^{*-}\bar{B}^{*0}\\
\dfrac{1}{\sqrt{2}}(B^{*0}\bar{B}^{*0}-B^{*+}B^{*-}),
\end{cases} \\
&I=0:&\dfrac{1}{\sqrt{2}}(B^{*0}\bar{B}^{*0}+B^{*+}B^{*-}).
\end{eqnarray}
For $D\bar{D}^*$ and $D^*\bar{D}^*$ systems, the similar expressions can be obtained.

\subsection{\textbf{$B\bar{B}^*$}}

By using the analytical effective interactions we have gotten in section II, we study the $B\bar{B}^*$ system
with different isospin $I$ and $C$-parity $C$. After solving the Schr\"{o}dinger Equation with the programs
developed in Refs. \cite{schoberl1985, lucha1999} and with the interactions
in Eq. \ref{eff:scalar},\ref{eff:pseudoscalar},\ref{eff:vector}, we list the obtained
results in Table \ref{BBstart}. In the table, I, II and III refer to the chiral SU(3) quark model, the extended
chiral SU(3) quark models with vector nonets coupling constant $f_{chv}/g_{chv}$=0 and 2/3, respectively.
If bound state does exist, we list, in the table, the binding energy $E$ in MeV. The symbol "---" denotes
that bound state doesn't exist.

\begin{table}[htbp]
\caption{Binding energies (MeV) of the bound states of $B\bar{B}^*$ system with different $I$ and $C$.
}\label{BBstart}
\centering
\begin{tabular*}{5cm}{@{\extracolsep{\fill}}ccccc}
\hline
$I$&$C$& I & II & III  \\
 \hline
\multirow{2}{*}{0}&+&46.74&97.75&78.04\\
&--&3.27&66.53&50.59\\  \hline
 \multirow{2}{*}{1}&+& --- & --- & ---\\
&--&0.39&1.75&1.49\\
 \hline
 \end{tabular*}
\end{table}

From Table \ref{BBstart}, we see there are three $S$-wave $B\bar{B}^*$ bound states
with quantum numbers $I^G(J^{PC})=0^{+}(1^{++})$, $0^{-}(1^{+-})$ and $1^{+}(1^{+-})$ respectively.
The binding energies of $0^{+}(1^{++})$, $0^{-}(1^{+-})$ and $1^{+}(1^{+-})$  bound states are respectively
46.76--97.75 MeV, 3.27--66.53 MeV and 0.39--1.75 MeV in our models. Table I shows that the attractive interaction
of I=0 $B\bar{B}^*$ system is much stronger than that of I=1 $B\bar{B}^*$ system and the attractive interaction
in the extended chiral SU(3) quark model (models II and III) is stronger than that in the chiral SU(3)
quark model (model I). From our calculation, we notice that in $B\bar{B}^*$ system, $\sigma$, $\sigma'$,
$\pi$, $\omega$, and $\rho$ exchange interactions play a dominant role and they determine
whether $B\bar{B}^*$ bound state exists or not.

For the $I$=0 $C$=+ case, in the chiral SU(3) quark model, $\sigma$, $\sigma'$ and $\pi$ exchanges provide
strong attraction, so the total interaction is strong enough to form a $B\bar{B}^*$ bound state.
In the extended chiral SU(3) quark model, the contributions of vector meson exchange are also included,
and $\rho$ and $\omega$ exchanges provide additional attraction. Therefore, the $B\bar{B}^*$ system has
a larger binding energy as shown in Table I where this $0^{+}(1^{++})$ bound state has a binding
energy 46.76--97.75 MeV.

For the $I$=0 $C$=-- case, in the chiral SU(3) quark model, $\sigma$ and $\sigma'$ exchanges provide
strong attraction but $\pi$ exchange provides strong repulsion, as a result of the competition,
the $B\bar{B}^*$  still binds but the binding energy becomes much smaller. In the extended chiral
SU(3) quark model, $\rho$ and $\omega$ exchanges provide very strong attraction as well, so the
binding energy becomes larger than the one in the chiral quark model. This $0^{-}(1^{+-})$
bound state has a binding energy 3.27--66.53 MeV.

For the $I$=1 $C$=+ case, in all the three quark models, $\sigma$ exchange provides strong attraction
but $\sigma'$ and $\pi$ exchanges provide strong repulsion, thus the total interaction is weakly attractive,
however the $B\bar{B}^*$  can't bind in our approach.

For the $I$=1 $C$=+ case, in the chiral SU(3) quark model, $\sigma$ exchange provides strong attraction
but $\sigma'$ and $\pi$ exchanges provide strong repulsion, thus the total interaction is weakly attractive,
however the $B\bar{B}^*$  can't bind in our approach. In the extended chiral SU(3) quark model,
the strong attraction provided by $\omega$ exchange and the strong repulsion provided by $\rho$ exchange
almost cancel each other, so the $B\bar{B}^*$  still can't bind.

\begin{figure}[htb]
\centering
\includegraphics[scale=0.3]{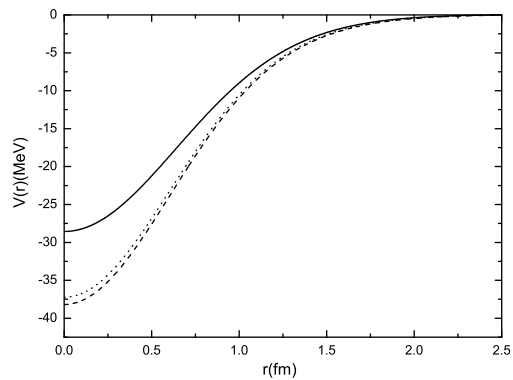}
\caption{The total interaction potential of $I$=1 and $C$=-- $B\bar{B}^*$ system.  The solid, dashed and dotted lines represent
the result obtained from the chiral SU(3) quark model, the extended chiral SU(3) quark model with $f_{chv}/g_{chv}$ taken as 0 and 2/3, respectively.  \label{BBstar1+1+-} }
\end{figure}

For the $I$=1 $C$=-- case, in the chiral SU(3) quark model, $\sigma$ and $\pi$ exchanges provide strong attraction
but $\sigma'$ exchange provides repulsion, so the total interaction is attractive,
and the $B\bar{B}^*$  can form a weakly bound state. In the extended chiral SU(3) quark model,
the contributions of vector meson exchange are also cancelled, so the $B\bar{B}^*$
still can form a $1^{+}(1^{+-})$ weakly bound state with weak binding energy 0.39--1.75 MeV.
In Fig. \ref{BBstar1+1+-}, the total interaction potential of the charged $1^{+}(1^{+})$ $S$-wave $B\bar{B}^*$
system is depicted. This charged $1^{+}(1^{+})$ $S$-wave $B\bar{B}^*$ weakly bound state has a mass 10603-10604 MeV,
and it might be explained as one possible candidate of $Z_b(10610)$.

\subsection{\textbf{$B^*\bar{B}^*$}}

By solving the Schr\"{o}dinger Equation with the analytical potentials of $B^*\bar{B}^*$ system
with different isospin $I$ and spin $S$, we list our results of the bound state problems of
the system in Table \ref{BstarBstart}.

\begin{table}[htbp]
\caption{Binding energies (MeV) of the bound states of $B^*\bar{B}^*$ system with different $I$ and $S$ quantum numbers.
Symbols have the same meaning as in Table \ref{BBstart}.}\label{BstarBstart}
\centering
\begin{tabular*}{5cm}{@{\extracolsep{\fill}}ccccc}
\hline
$I$&$S$& I & II & III  \\
 \hline
\multirow{3}{*}{0}&0& --- &51.65&37.55\\
&1& 3.30&66.73&50.77\\
&2& 46.91 &97.98&78.24\\  \hline
\multirow{3}{*}{1}&0&1.51&3.66&3.32\\
&1& 0.41&1.78&1.52\\
&2& --- & --- & ---\\  \hline
 \end{tabular*}
\end{table}

According to Table \ref{BstarBstart}, we see that four $S$-wave bound states with different quantum numbers
could exist in $B^*\bar{B}^*$ system, namely, the bound states of $1^+(1^{+-})$, with binding energy
0.41--1.78 MeV, $0^-(1^{+-})$ with binding energy 3.30--66.73 MeV, $0^+(2^{++})$ with binding energy
46.91--97.98 MeV and $1^-(0^{++})$ with binding energy 1.51--3.66 MeV. From our calculation, we notice
that, in $B^*\bar{B}^*$ system, $\sigma$, $\sigma'$, $\pi$, $\omega$ and $\rho$  exchanges also play a
dominant role to determine whether $B^*\bar{B}^*$ could form a bound state or not.

For the $I$=0 $S$=0 case, in the chiral SU(3) quark model (model I), $\sigma$ and $\sigma'$ exchanges
provide attraction, however, $\pi$ exchange provides strong repulsion. As a result of this competition,
the total interaction is too weak to bind  $B^*\bar{B}^*$. In the extended chiral SU(3) quark model
(models II and III), because the contributions of vector meson exchange are also included, and
$\rho$ and $\omega$ exchanges also provide very strong attraction, the $B^*\bar{B}^*$ can form a bound state.

For the $I$=0 $S$=1 case, in the chiral SU(3) quark model, $\sigma$ and $\sigma'$ exchanges provide attraction
but $\pi$ exchange provides repulsion, and the total interaction in the $B^*\bar{B}^*$ system is still strong
enough to form a bound state.  In the extended chiral SU(3) quark model, the additional $\rho$ and $\omega$
exchanges provide very strong attraction, therefore, the binding energy becomes larger. This $0^{-}(1^{+-})$
bound state has a binding energy of 3.30--66.73 MeV.

For the $I$=0 $S$=2 case, in the chiral SU(3) quark model, all $\sigma$, $\sigma'$ and $\pi$ exchanges provide
strong attraction, so the $B^*\bar{B}^*$  forms a bound state. Moreover, in the extended chiral SU(3) quark model,
$\rho$ and $\omega$ exchanges provide additional attraction, so the $B^*\bar{B}^*$ system could be deeply bound.
This $0^{+}(2^{++})$ bound state has a binding energy 46.91--97.98 MeV.

\begin{figure}[htb]
\centering
\includegraphics[scale=0.3]{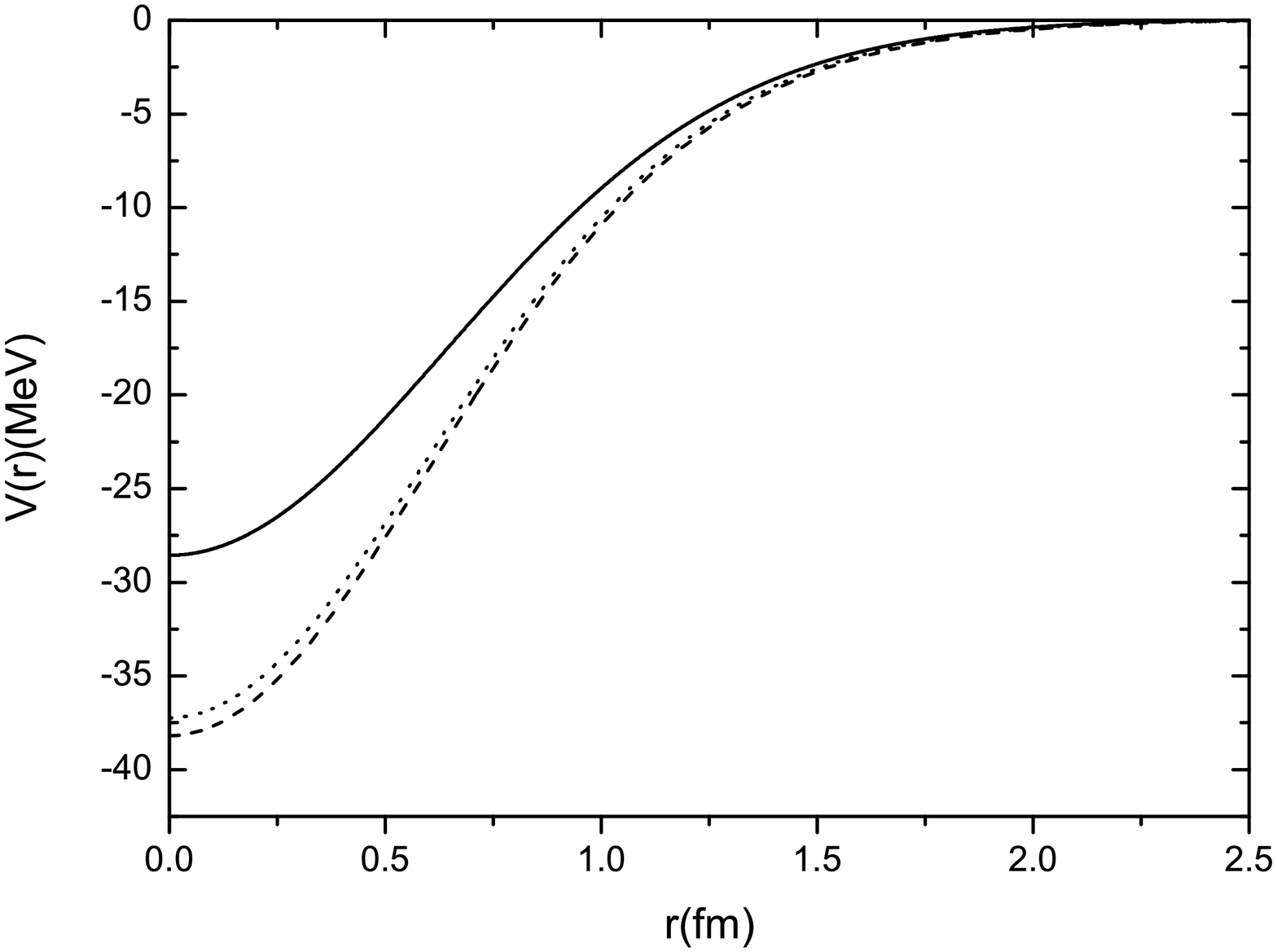}
\caption{The total interaction potential of $I$=1 $S$=1 $B^*\bar{B}^*$ system.
Symbols have the same meaning as in Fig. \ref{BBstar1+1+-}.
\label{BstarBstar1+1+-} }
\end{figure}

For the two cases of $I$=1 $S$=0 and 1, in the chiral SU(3) quark model, $\sigma$ and $\pi$ exchanges provide
attraction, and $\sigma'$ exchange provides repulsion. However, the total interaction is attractive,
and the $B^*\bar{B}^*$  can form a weakly bound state. In the extended chiral SU(3) quark model,
like the $I$=1 $B\bar{B}^*$ system, the contributions of vector meson exchange are also
almost cancelled, and the $B^*\bar{B}^*$  still can form weakly bound states.
The $1^{-}(0^{++})$ and $1^{+}(1^{+-})$ bound states have weakly binding energies 1.51--3.66 MeV and
0.41--1.78 MeV, respectively. In Fig. \ref{BstarBstar1+1+-}, the total interaction potential of the
charged $1^{+}(1^{+})$ $S$-wave $B^*\bar{B}^*$ system is plotted. It should be stressed that
this $1^{+}(1^{+})$ $B^*\bar{B}^*$ weakly bound state has a mass 10648-10650 MeV,
which is consistent with the mass and quantum numbers of $Z_b(10650)$. As a result, the new resonance
of $Z_b(10650)$ might be explained as the weakly bound state of $1^{+}(1^{+})$ $B^*\bar{B}^*$ in our approach.

Comparing Fig. \ref{BstarBstar1+1+-} with Fig. \ref{BBstar1+1+-}, we see the two interaction potentials are
almost the same. It is not surprising because in these two cases, the spin matrix elements in
Eqs. \ref{eff:pseudoscalar} and \ref{eff:vector} are both --1 and the other parts of the
interaction potentials are the same as each other. The $B^*\bar{B}^*$ system has a slightly larger binding
energy because of its larger reduced mass than the $B\bar{B}^*$ system which is introduced when solving
the Schrodinger Equation.

For the $I$=1 $S$=2 case of $B^*\bar{B}^*$ system, we found, in all the three SU(3) quark models,
$\sigma$ exchange provides attraction but $\sigma'$ and $\pi$ exchanges provide repulsion, and the total
attractive interaction is too weak to bind the $B^*\bar{B}^*$.

\subsection{\textbf{$D\bar{D}^*$}}

The only difference between the systems of $D\bar{D}^*$ and $B\bar{B}^*$ is the flavor of heavy quark .
Considering the resemblance between $D\bar{D}^*$ and $B\bar{B}^*$, the one meson exchange interaction
potentials should have similar properties. In our calculation we find the one meson exchange interaction
potentials of $D\bar{D}^*$ system are a bit larger than those of $B\bar{B}^*$ system with the same quantum
numbers of isospin $I$ and $C$-parity $C$, and the total interaction also is a bit larger.
That is to say, the heavy quark mass has only a little effect on the interaction potentials (see Eq. 19).
After solving the Schr\"{o}dinger Equation, we list our results in Table \ref{DDstart}.

\begin{table}[htbp]
\caption{Binding energies (MeV) of the bound states of $D\bar{D}^*$ system with different $I$ and $C$
quantum numbers. Symbols have the same meaning as in Table \ref{BBstart}.}\label{DDstart}
\centering
\begin{tabular*}{5cm}{@{\extracolsep{\fill}}ccccc}
\hline
$I$&$C$& I & II & III  \\
 \hline
\multirow{2}{*}{0}&+&12.72&47.39&33.32\\
&--&--- &24.26&14.24\\  \hline
 \multirow{2}{*}{1}&+& --- & --- & ---\\
&--& --- & --- & ---\\
 \hline
 \end{tabular*}
\end{table}

From Table \ref{DDstart}, we find the result of $D\bar{D}^*$ system is quite different from that
of $B\bar{B}^*$ system. This is because the much lighter reduced mass makes the $D\bar{D}^*$ system more difficult
to form a bound state. We see that only one $S$-wave $0^{+}(1^{++})$ $D\bar{D}^*$ bound state
with binding energy of 12.72--47.39 MeV could exist. Unlike the $B\bar{B}^*$ system,
the total attractive interactions are too weak to form $0^-(1^{+-})$ and $1^+(1^{+-})$ $D\bar{D}^*$
states.

\begin{figure}[htb]
\centering
\includegraphics[scale=0.3]{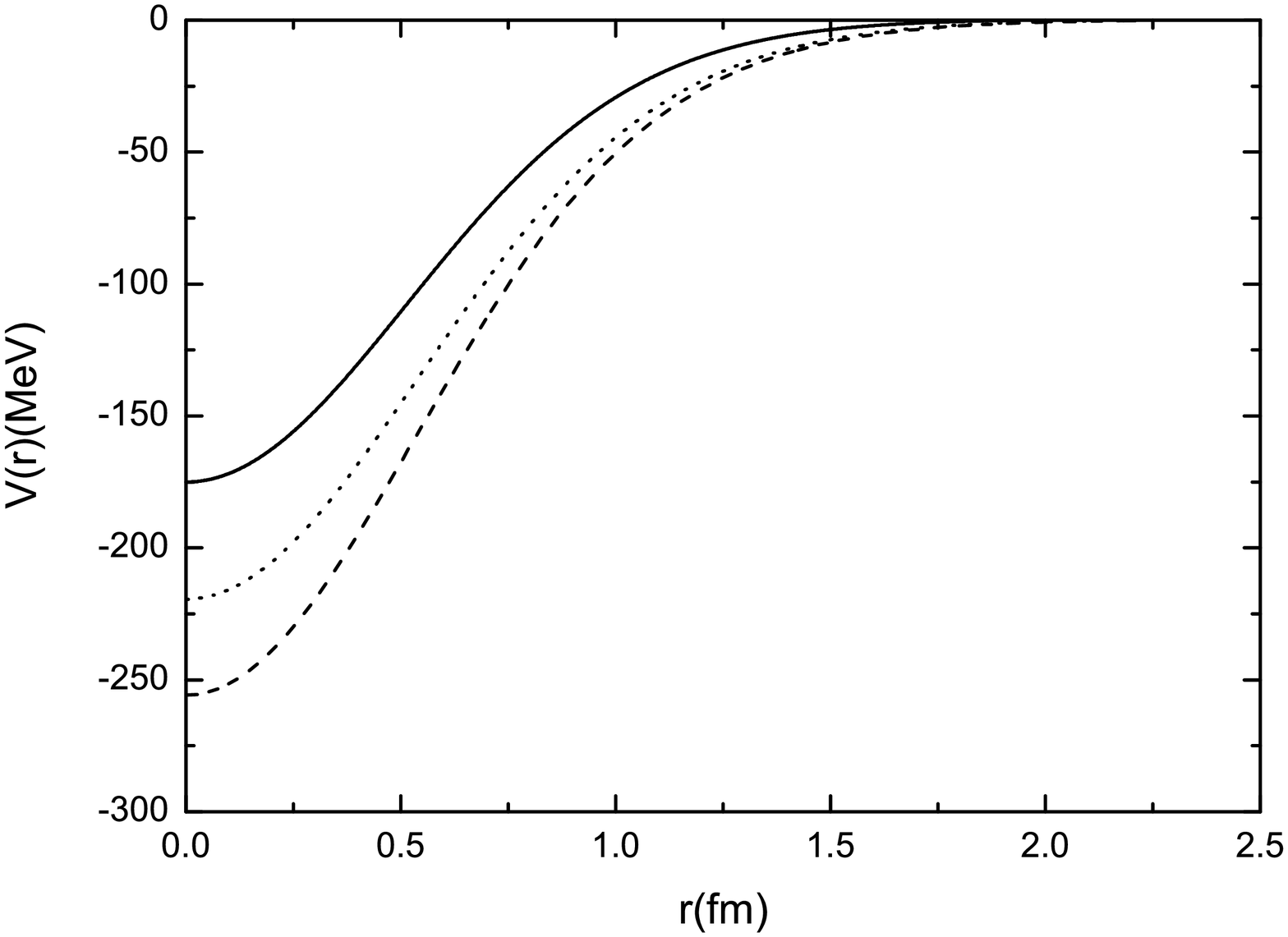}
\caption{The total interaction potential of $I$=0 $C$=+ $D\bar{D}^*$ system.
Symbols have the same meaning as in Fig. \ref{BBstar1+1+-}. \label{DDstar0+1++} }
\end{figure}

In Fig. \ref{DDstar0+1++}, the total interaction of $I$=0 $C$=+ $D\bar{D}^*$ system is shown.
Our $0^{+}(1^{++})$ $D\bar{D}^*$ bound state has a mass 3832-3867 MeV,
and it fits well with the mass of new resonance of $X(3872)$ observed by \cite{belle1, cdf, d0, barbar0}.
Particularly, the quantum numbers of $0^{+}(1^{++})$ seem to be more preferable in the recent
experiment \cite{jens} of X(3872). Here we speculate that the study of the $0^{+}(1^{++})$ $D\bar{D}^*$
molecule bound state may shed light on the $X(3872)$ structure.

\subsection{\textbf{$D^*\bar{D}^*$}}

In $D^*\bar{D}^*$ system the one meson exchange interactions have similar properties to
those of $B^*\bar{B}^*$ system. Analogous to the process of $B^*\bar{B}^*$ system, we list
the result of $D^*\bar{D}^*$ system in Table \ref{DstarDstart}.
\begin{table}[htbp]
\caption{Binding energies (MeV) of the bound states of $D^*\bar{D}^*$
system with different isospin $I$ and spin $S$.
Symbols have the same meaning as in Table \ref{BBstart}.}\label{DstarDstart}
\centering
\begin{tabular*}{5cm}{@{\extracolsep{\fill}}ccccc}
\hline
$I$&$S$& I & II & III  \\
 \hline
\multirow{3}{*}{0}&0& --- &16.15&7.97\\
&1& --- &26.38&15.89\\
&2&14.33&50.22&35.69\\  \hline
\multirow{3}{*}{1}&0 & --- & --- & --- \\
&1& --- & --- & --- \\
&2& --- & --- & ---\\  \hline
 \end{tabular*}
\end{table}

Comparing Table \ref{DstarDstart} with Table \ref{BstarBstart}, we find the $D^*\bar{D}^*$ system is more difficult
to form bound states than the $B^*\bar{B}^*$ system. This is because the reduced mass of the
$D^*\bar{D}^*$ system is 1003 MeV which is lighter than that of $B^*\bar{B}^*$ system of 2663 MeV.
For $D^*\bar{D}^*$ system, we see there is only one $0^+(2^{++})$  $D^*\bar{D}^*$ bound state with
binding energy of 14.33--50.22 MeV. Unlike the $B^*\bar{B}^*$ system, the total attractive interactions
are too weak to form $0^-(1^{+-})$, $1^-(0^{++})$ and $1^+(1^{+-})$ $D^*\bar{D}^*$ states.

\begin{figure}[htb]
\centering
\includegraphics[scale=0.3]{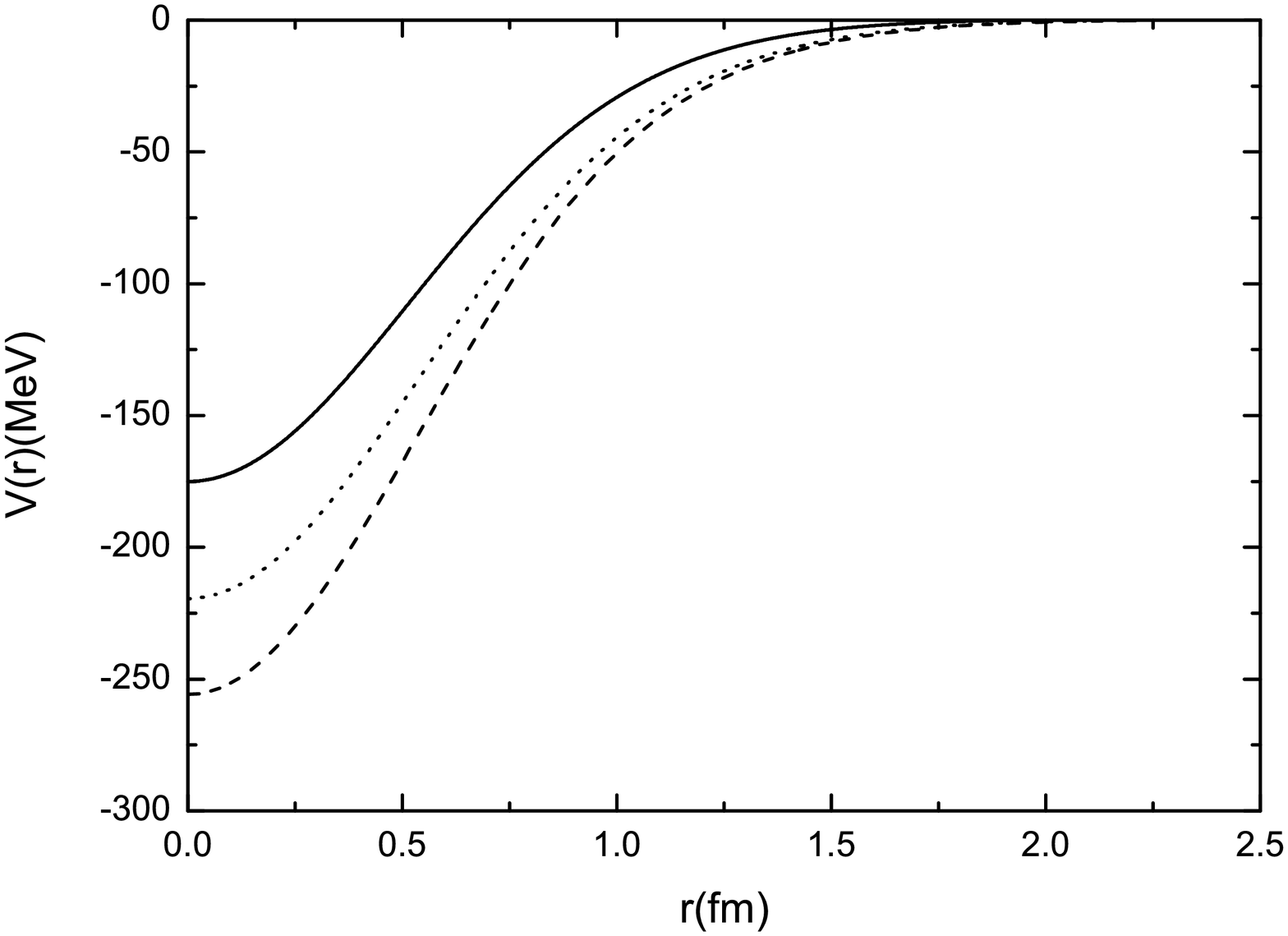}
\caption{The total interaction potential of $I$=0 $S$=2 $D^*\bar{D}^*$ system. Symbols have the same meaning as in Fig. \ref{BBstar1+1+-}. \label{DstarDstar0+2++} }
\end{figure}

The total interaction potential of $I$=0 $S$=2  $D^*\bar{D}^*$ system is depicted in Fig. \ref{DstarDstar0+2++}.
This $0^{+}(2^{++})$ $D^*\bar{D}^*$ state has a mass of 3964-3999 MeV. We notice that
the mass and quantum numbers of this state are consistent with those of $Y(3940)$ observed
recently  by Refs. \cite{belle2005,belle2006,belle2007,barbar2008} and discussed by
Refs. \cite{oset,stephen}. Therefore, the study of our $0^{+}(2^{++})$ $D^*\bar{D}^*$ molecule
state might be helpful to understand the structure of the $Y(3940)$.

\section{summary}\label{sec:sum}

In this work, we have performed a systematic investigation of the bound state problem of
$D\bar{D}^*,D^*\bar{D}^*$, $B\bar{B}^*$ and $B^*\bar{B}^*$ systems.
In $D\bar{D}^*$ system, we find only one $0^+(1^{++})$ bound state
and this molecule state might partly account for the structure of $X(3872)$.
In $D^*\bar{D}^*$ system, we find only one $0^+(2^{++})$  bound state could exist
and this $D^*\bar{D}^*$ molecule state might be explained as $Y(3940)$ observed
in experiments.

General speaking, we found that $B\bar{B}^*$ and $B^*\bar{B}^*$ can form bound states
more easily than $D\bar{D}^*$ and $D^*\bar{D}^*$, because of their much heavier reduced masses.
For the two charged new resonances of $Z_b(10610)$ and $Z_b(10650)$, our calculation shows
that they might be explained as $1^+(1^{+})$ weakly bound states of $B\bar{B}^*$ and
$B^*\bar{B}^*$, respectively. For $B\bar{B}^*$ system, other two bound states are found
in our approach as  $0^+(1^{++})$ with a mass 10506--10557 MeV and $0^-(1^{+-})$ with
a mass 10538--10600 MeV. For $B^*\bar{B}^*$ system, other three bound states are also found.
They are $0^+(2^{++})$ with a mass 10552--10603 MeV, $0^-(1^{+-})$ with a mass 10583--10645 MeV,
and $1^+(0^{+-})$ with a mass 10646--10649 MeV. However, until now no experimental evidence
has been reported for these five bound states in our approach.

What's more, there are some states unbound in our chiral SU(3) quark model (model I) but bound in
our extended chiral SU(3) quark model (model II and model III). So far, we can't draw definite
conclusions for them. It is expected future experimental measurements will provide more
information about these new hadron structures and provide a discrimination for
our model calculations.

\begin{acknowledgments}
This project was supported in part by National Natural Science Foundation of China (Grant Nos. 11105158,
11035006,10975146), Ministry of Science and Technology of China (Grant No. 2009CB825200), and China
Postdoctoral Science Foundation (Grant No. 20100480468).
\end{acknowledgments}

\end{document}